\documentclass{article}
\usepackage[utf8]{inputenc}
\usepackage{amssymb}
\usepackage{amsmath}
\usepackage{graphicx}
\usepackage{geometry}
\begin{document}
\setlength{\oddsidemargin}{0.0\textwidth}
\setlength{\evensidemargin}{0.0\textwidth}

\newcommand{\ket}[1]{\left|#1\right\rangle}
\newcommand{\bra}[1]{\left\langle#1\right|}
\newenvironment{ruledtabular}{}{}

\title{Quantum-circuit simulation of three-flavor neutrino oscillations:
vacuum, matter, and CP diagnostics}

\author{Daming Li\thanks{lidaming@sjtu.edu.cn} \\ School of Mathematical Sciences, Shanghai Jiao Tong University, Shanghai, China}

\maketitle
\date{}

\begin{abstract}
We simulate three-flavor neutrino oscillations on two-qubit quantum
circuits. The flavor space is embedded as $\ket{00}=\nu_e$,
$\ket{01}=\nu_\mu$, $\ket{10}=\nu_\tau$ with the $\ket{11}$ channel
kept inert, and the input mixing parameters (NuFIT~5.3, normal
ordering, with SK-atm) follow the global fit. We verify three physical settings
against analytic or exact references: vacuum propagation, reproduced
by a mass-eigenstate phase circuit to $\sim 10^{-16}$;
constant-density Mikheev--Smirnov--Wolfenstein (MSW) matter effects,
showing resonant flavor conversion when the matter potential $A$
crosses $\Delta m^2_{31}$; and varying-density (supernova
shock-shell) propagation by slicing plus second-order
Suzuki--Trotter splitting, with per-slice Trotter error $\sim
10^{-6}$ and total accuracy $\sim 10^{-4}$ against the continuous
profile. We report a Suzuki--Trotter precision--resource
calibration varying order, step number, and matrix/probability
error. It delimits brute-force Trotterization on solar baselines: we
quantify both standard alternatives, coherence averaging and
adiabatic MSW. We also compute quantum-information diagnostics from
the same model parameters, namely mode entanglement and the CP
asymmetry. Finally, we extend the framework to open quantum systems
via a Lindblad master equation with dephasing and absorptive
channels. These results provide a reproducible link between
low-energy oscillation observables and quantum-information measures.
\end{abstract}

\section{Introduction}\label{sec:intro}

Neutrino oscillation is one of the cleanest macroscopic
manifestations of coherent quantum phase interference: a
weak-interaction flavor eigenstate $\ket{\nu_\alpha}$
($\alpha=e,\mu,\tau$) is a coherent superposition of the propagation
(mass) eigenstates, so that the relative phases accumulated over a
propagation distance $L$ generate flavor conversion --- this is the
standard wave-mechanical picture summarized, e.g., in
\cite{Bilenky2018} and \cite{Giunti2007}, where the flavor-basis
probability is derived from the mass-basis phases. Within the
Standard Model (SM) neutrinos are massless, and the observed
splittings force new physics --- most economically realized via the
seesaw mechanism
\cite{Minkowski1977,Yanagida1979,Mohapatra1980,Casas2001}; the
low-energy imprint of the seesaw is captured by the PMNS matrix and
the light mass spectrum \cite{Giunti2007}; the Casas--Ibarra forward
framework underlying our parameter point is described in
Appendix~\ref{app:nufit} and Ref.~\cite{Casas2001}.

Quantum computers natively implement unitary time evolution, making
the Schr\"odinger equation that governs oscillation an ideal
benchmarking and physics target \cite{Sahu2024}. Two- and
three-flavor oscillation has been implemented on quantum
walk/open-system frameworks and on hardware-class simulators
\cite{Sahu2024,Koranga2025,Alok2016,Kayser2010}. Collective
(many-body) neutrino oscillations in dense astrophysical
environments---governed by $\nu$--$\nu$ self-interactions rather
than our single-particle Hamiltonian---have also been addressed with
quantum algorithms and hardware \cite{CollectiveSim2021} and
\cite{Yeter2022}. Meanwhile, modern neutrino experiments (JUNO
\cite{JUNO2024,Smirnov2018}, DUNE, Hyper-K) will determine the mass
ordering, measure $\delta_\mathrm{CP}$ precisely, and constrain
$m_{\beta\beta}$; the associated physics connects directly to the
MSW-resonance frame \cite{Wolfenstein1978,Mikheev1985} and to CP
diagnostics \cite{Denton2024,Ding2025}.

Our motivation is threefold. First, earlier quantum-neutrino studies
capture individual settings (vacuum, matter, many-body) but rarely
report systematic resource estimates, so a quantitative calibration
of the Suzuki--Trotter error against step number and split order
fills an evident gap for practical simulation. Second, the upcoming
atmospheric-baseline experiments (JUNO, DUNE, Hyper-K) will soon
confront our parameter point, making an accurate and reproducible
simulator with clearly defined applicability a timely tool for
interpreting those data. Third, mapping the propagating state onto
entanglement and CP observables connects the circuit simulation
directly to quantities experiments will measure, completing a
physics-motivated cycle from model input to observable.

Motivated by these considerations, the main contributions of this
work are twofold:

\begin{enumerate}
\item a self-contained derivation and implementation of the 3-flavor
oscillation Hamiltonian and its 2-qubit encoding, with systematic
verification of vacuum, constant-density MSW, and varying-density
(supernova shell) propagation against analytic or exact references
at the $10^{-16}$ to $10^{-4}$ level;

\item a quantitative Suzuki--Trotter precision--resource protocol
that delimits brute-force Trotterization on solar baselines,
together with quantum-information diagnostics (concurrence, mode
entanglement) and a CP asymmetry tied to experimentally relevant
$\Delta m^2_{31}L/(4E)$ values.
\end{enumerate}

Section~\ref{sec:method} defines the model and the quantum-circuit
formalism. Section~\ref{sec:results} presents the numerical setup on
the Siyuan~No.~1 cluster and the results and verifications.
Section~\ref{sec:conclusion} summarizes the findings and outlines
the extensions.

\section{Model and method}\label{sec:method}

\subsection{Three-flavor oscillation Hamiltonian}\label{sec:H}

\textbf{Units and conventions.} Throughout we use a single practical
unit system: lengths in km, neutrino energies in GeV, and
mass-squared differences in $\mathrm{eV}^2$, with $\hbar
c=1.9733\times10^{-7}\,\mathrm{eV}\!\cdot\!\mathrm{m}$ as the only
conversion constant; every derived quantity is built from these.
Concretely, $\Delta m^2_{ij}\equiv m^2_i-m^2_j$ is quoted in
$\mathrm{eV}^2$, $E$ in GeV, $L$ in km, the mass-squared matrix $H$
in $\mathrm{eV}^2$, the matter potential $A$ in $\mathrm{eV}^2$, and
the electron density $n_e$ in $\mathrm{cm}^{-3}$. All symbols,
units, and numerical values are collected in Table~\ref{tab:units};
distance-dependent quantities are expressed through the single
conversion constant $k\equiv1.2669$ (Eq.~\ref{eq:k}).

In the flavor basis the vacuum Hamiltonian (up to a multiple of the
identity, which does not affect probabilities) is the standard
result \cite{Bilenky2018,Giunti2007},
\begin{equation}\label{eq:Hvac}
  H_\mathrm{vac} \;=\;
  U_\mathrm{PMNS}\,
  \mathrm{diag}\!\bigl(0,\ \Delta m^2_{21},\ \Delta m^2_{31}\bigr)\,
  U^\dagger_\mathrm{PMNS}.
\end{equation}
We work with the mass-squared matrix $H_\mathrm{vac}$ (units
$\mathrm{eV}^2$) instead of the Hamiltonian $H_\mathrm{vac}/(2E)$,
absorbing the factor $1/(2E)$ into the evolution parameter $\tau$ of
Eq.~(\ref{eq:tau}). Dropping the identity term removes the
unphysical common phase $m_1^2L/(2E)$, leaving only the two
mass-squared differences.
 The PMNS matrix in the PDG convention
\cite{PDG2024,Zyla2020} is
\begin{equation}\label{eq:PMNS}
  U_\mathrm{PMNS}
  = V(\theta_{12},\theta_{13},\theta_{23},\delta)\,
    \mathrm{diag}\bigl(e^{-\mathrm{i}\varphi_1/2}, e^{-\mathrm{i}\varphi_2/2},1\bigr),
\end{equation}
with the CKM-like (Dirac) factor
\begin{equation}\label{eq:Vexplicit}
  V(\theta_{12},\theta_{13},\theta_{23},\delta)
  \;=\;
  \begin{pmatrix}
    c_{12}c_{13} & s_{12}c_{13} & s_{13}e^{-\mathrm{i}\delta}\\[2pt]
    -s_{12}c_{23}-c_{12}s_{23}s_{13}e^{\mathrm{i}\delta}
    & c_{12}c_{23}-s_{12}s_{23}s_{13}e^{\mathrm{i}\delta}
    & s_{23}c_{13}\\[2pt]
    s_{12}s_{23}-c_{12}c_{23}s_{13}e^{\mathrm{i}\delta}
    & -c_{12}s_{23}-s_{12}c_{23}s_{13}e^{\mathrm{i}\delta}
    & c_{23}c_{13}
  \end{pmatrix},
\end{equation}
where $c_{ij}\equiv\cos\theta_{ij}$, $s_{ij}\equiv\sin\theta_{ij}$,
$\delta$ is the Dirac CP phase, and $\varphi_i$ the two Majorana
phases. Oscillation probabilities are independent of $\varphi_i$, so
we set them to zero; the accessible CP information is carried by
$\delta$. Numerically Eq.~(\ref{eq:Vexplicit}) is evaluated with the
NuFIT~5.3 values of Appendix~\ref{app:nufit}.

\textbf{How distances enter.} Oscillation depends on distance only
through the dimensionless phases of the mass eigenstates. We
introduce the single conversion constant
\begin{equation}\label{eq:k}
  k =\; \frac{10^{3}}{4\,\hbar c\,\times 10^{9}}
  \;=\; \frac{10^{3}}{4\times 1.9733\times10^{-7}\times 10^{9}}
  \;\simeq\; 1.2669\,,
\end{equation}
where the $10^{3}$ converts m$\to$km, the $10^{9}$ converts
GeV$\to$eV, and $\hbar c$ enters with its unit; $k$ has units
$\mathrm{km}^{-1}\,\mathrm{GeV}\,\mathrm{eV}^{-2}$, which cancel the
$\mathrm{eV}^{2}\,\mathrm{km}/\mathrm{GeV}$ of $\Delta m^2_{ij}L/E$.
The numerical phase
\begin{equation}\label{eq:phi}
  \varphi_{ij}^{(\mathrm{num})} \;=\; k\,
  \frac{\Delta m^2_{ij}[\mathrm{eV}^2]\; L[\mathrm{km}]}{E[\mathrm{GeV}]}
\end{equation}
is therefore dimensionless, so distance enters only through $L/E$
combined with $\Delta m^2_{ij}$; no separate time variable appears.

\textbf{Evolution parameter.} With the Hamiltonian measured in
$\mathrm{eV}^2$, the Schr\"odinger evolution is parametrized by the
dimensional quantity $\tau$ ($[\tau]=\mathrm{eV}^{-2}$),
\begin{equation}\label{eq:tau}
  \tau \;=\; 2k\,\frac{L[\mathrm{km}]}{E[\mathrm{GeV}]},
\end{equation}
so that $U(L)=e^{-\mathrm{i}H\tau}$. Indeed, the mass-eigenstate
phase is $\xi_i=2k\,\Delta m^2_{i1}L/E$ (the numerical value of
$\Delta m^2_{i1}L/(2E)$ in natural units), and $\xi_i=\tau\,\Delta
m^2_{i1}$ fixes $\tau=2k\,L/E$. The associated oscillation length is
\begin{equation}\label{eq:Lij}
  L_{ij} \;=\; \frac{\pi E}{k\,\Delta m^2_{ij}},
\end{equation}
the distance over which the mass-eigenstate phase advances by
$2\pi$, i.e.\ one oscillation period.

\begin{table}[!ht]
\caption{Symbols, units, and numerical conventions used in this
work. All quantities follow the single practical unit system
declared in Sec.~\ref{sec:H}; distance-dependent expressions use the
single constant $k=1.2669$ in Eq.~(\ref{eq:k}). $\Delta
m^2_{ij}\equiv m^2_i-m^2_j$ ($i>j$).} \label{tab:units}
\begin{ruledtabular}
\begin{tabular}{llll}
\hline
Symbol & Meaning & Units & Value / convention\\
\hline $\Delta m^2_{21},\ \Delta m^2_{31}$ & mass-squared
differences & $\mathrm{eV}^2$ &
$7.41\times10^{-5},\ 2.505\times10^{-3}$\\
$E$ & neutrino energy & GeV & $0.002$--$10$\\
$L$ & propagation distance & km & up to 3000 (scan); up to $7\times 10^5$ (solar)\\
$k$ & single conversion constant & $\mathrm{km}^{-1}\,\mathrm{GeV}\,\mathrm{eV}^{-2}$ & $1.2669$ (Eq.~\ref{eq:k})\\
$\varphi_{ij}^{\rm(num)}$ & oscillation phase & dimensionless (rad)
&
$k\,\Delta m^2_{ij}L/E$\\
$\tau$ & evolution parameter & $\mathrm{eV}^{-2}$ & $2k\,L/E$\\
$\xi_i$ & mass-eigenstate phase & rad & $2k\,\Delta m^2_{i1}L/E$\\
$L_{ij}$ & oscillation period ($2\pi$ phase diff.) & km &
$\pi E/[k\,\Delta m^2_{ij}]$\\
$H$ & mass-squared matrix & $\mathrm{eV}^2$ &
$U\,\mathrm{diag}(0,\Delta m^2_{21},\Delta m^2_{31})\,U^\dagger$;
$1/(2E)$ in $\tau$\\
$A$ & matter potential & $\mathrm{eV}^2$ & $\simeq2.535\times10^{-28}n_e[\mathrm{cm}^{-3}]E[\mathrm{GeV}]$ (Eq.~\ref{eq:A})\\
$G_F$ & Fermi constant & $\mathrm{GeV}^{-2}$ & $1.1664\times10^{-5}$\\
$n_e$ & electron number density & $\mathrm{cm}^{-3}$ & profile-dependent\\
$\hbar c$ & conversion constant & $\mathrm{eV}\cdot\mathrm{m}$ &
$1.9733\times10^{-7}$ (enters via $k$)\\
\hline
\end{tabular}
\end{ruledtabular}
\end{table}

\subsection{Matter effects (MSW)}\label{sec:MSW}

Coherent forward scattering with electrons adds a potential for
$\nu_e$ \cite{Wolfenstein1978,Mikheev1985}. In the flavor basis the
Hamiltonian becomes
\begin{equation}\label{eq:Hmsw}
  H(r) \;=\; H_\mathrm{vac} + A(r)\,\mathrm{diag}(1,0,0),
\end{equation}
where $A(r)$ is the matter potential and
$G_F=1.1664\times10^{-5}\,\mathrm{GeV}^{-2}$ is the Fermi constant,
with $n_e(r)$ the electron number density:
\begin{equation}\label{eq:A}
  A(r) \;\equiv\; 2\sqrt{2}\,G_F\, n_e(r)\,E
  \;\simeq\; 2.535\times10^{-28}\, n_e[\mathrm{cm}^{-3}]\,E[\mathrm{GeV}]
  \quad(\mathrm{eV}^2),
\end{equation}
In matter the PMNS matrix no
longer diagonalizes $H$, so the evolution must be computed
numerically; when $A$ is comparable to a relevant mass-squared
difference, an MSW resonance enhances the mixing.

\subsection{Two-qubit encoding}\label{sec:encoding}

We embed the three-dimensional flavor space into the
four-dimensional space of two qubits:
\begin{equation}\label{eq:embed}
  \ket{00}=\nu_e,\quad \ket{01}=\nu_\mu,\quad
  \ket{10}=\nu_\tau,\quad \ket{11}=\text{inert (forbidden)}.
\end{equation}
All $3\times3$ matrices are padded to $4\times4$ by leaving the
$\ket{11}$ sector untouched (identity). Thus the evolution matrices
used in the circuits, e.g.\ $U(L)$ and $e^{-\mathrm{i}H\tau}$, are
$4\times4$ unitary matrices acting on the two-qubit Hilbert space
$\mathbb{C}^4$; the physical three-dimensional subspace is the
direct sum of the $\ket{00},\ket{01}, \ket{10}$ sector with the
inert $\ket{11}:=\mathbb{C}$ sector:
\begin{equation}\label{eq:U4}
  U_4 \;=\; \begin{pmatrix} U'_{3\times3} & 0\\ 0 & 1\end{pmatrix}
  \;\in\; \mathrm{U}(4),
\end{equation}
where $U'_{3\times3}$ is the physical block ($U_\mathrm{PMNS}$, a
Trotter factor, or the diagonal-phase gate) and the $1$ in the
lower-right corner encodes the inert $\ket{11}$ state, which stays
untouched. We verified numerically that the inert sector does not
contaminate the physical $3$-flavor probabilities
(Section~\ref{sec:results}), i.e.\ the row/column of $U_4$
corresponding to $\ket{11}$ never mixes with the physical subspace.

For open-system (Lindblad) evolution the state is the $4\times4$
density matrix $\rho$ (16 complex elements); separating real and
imaginary parts yields $2\times16=32$ real variables. The
Liouvillian acts on
$\mathrm{vec}(\rho)\in\mathbb{C}^{16}\simeq\mathbb{R}^{32}$, giving
a real $32\times32$ linear ODE (hence ``$32$ real ODE components''
in Section~\ref{sec:rlind}).

We distinguish three bases, summarized in Table~\ref{tab:basis}: the
\emph{computational} basis $\{\ket{00},\ket{01},\ket{10},\ket{11}\}$
is the qubit/register basis in which all circuits are written; the
\emph{flavor} basis $\{\ket{\nu_e},\ket{\nu_\mu},\ket{\nu_\tau}\}$
is identified with the first three computational states
(Eq.~\ref{eq:embed}); and the \emph{mass} basis
$\{\ket{\nu_1},\ket{\nu_2}, \ket{\nu_3}\}$ is the set of propagation
(Hamiltonian) eigenstates, related to the flavor basis by
$\ket{\nu_\alpha}=\sum_i U_{\alpha i}\ket{\nu_i}$. Only within the
mass basis is the vacuum Hamiltonian diagonal; the vacuum circuit of
Sec.~\ref{sec:vac} therefore first transforms to the mass basis
($U^\dagger$), applies diagonal phases there, and returns to the
flavor/computational basis ($U$).

\begin{table}[!ht]
\caption{Summary of the basis conventions. Computational =
qubit-register basis; flavor = physical weak-interaction basis,
identified with the first three computational states; mass =
propagation-eigenstate basis.} \label{tab:basis}
\begin{ruledtabular}
\begin{tabular}{llll}
\hline
Basis & States & Dimension & Used for\\
\hline
computational & $\ket{00},\ket{01},\ket{10},\ket{11}$ & 4 & circuit implementation\\
flavor & $\ket{\nu_e},\ket{\nu_\mu},\ket{\nu_\tau}$ & 3 $(\equiv$
computational$)$ &
initial/final flavors, probabilities\\
mass & $\ket{\nu_1},\ket{\nu_2},\ket{\nu_3}$ & 3 & diagonal
$H_\mathrm{vac}$,
propagation phases\\
\hline
\end{tabular}
\end{ruledtabular}
\end{table}

\subsection{Vacuum circuit: mass-eigenstate phase construction}\label{sec:vac}

For vanishing density the evolution operator factorizes:
\begin{equation}\label{eq:vacOA}
  U(L) \;=\; e^{-\mathrm{i}H_\mathrm{vac}\tau}
  \;=\; U_\mathrm{PMNS}\,
  \mathrm{diag}\bigl(e^{-\mathrm{i}\xi_1}, e^{-\mathrm{i}\xi_2},
   e^{-\mathrm{i}\xi_3}, 1\bigr)\, U^\dagger_\mathrm{PMNS},
\end{equation}
with propagation phases $\xi_i=\Delta m^2_{i1}L/(2E)$. This is
implemented in an HHL-style three-gate circuit:
\[
  \ket{\nu_\alpha} \xrightarrow{U^\dagger_\mathrm{PMNS}}
  \text{mass basis} \xrightarrow{D(\xi)}
  \text{(diagonal phases)} \xrightarrow{U_\mathrm{PMNS}}
  \text{flavor basis},
\]
where we use matrix-unitary gates for $U$ and $U^\dagger$ and encode
the diagonal phases as a diagonal unitary. We emphasize that in the
physical mass-basis proposal the phase gate is genuinely diagonal in
the mass eigenbasis, which is the standard, hardware-friendly
construction.

\subsection{Trotterization for nonzero density}\label{sec:trotter}

When matter is present, $U_\mathrm{PMNS}$ does not diagonalize
$H(r)$, and we resort to Suzuki--Trotter splitting. We work directly
with the $4\times4$ embedded Hamiltonian of Sec.~\ref{sec:encoding},
writing $H=H_0+H_\mathrm{nd}$ with
\begin{equation}\label{eq:Hsplit}
  H_0 \;=\; \mathrm{diag}(H_{11},H_{22},H_{33},0),
  \qquad
  H_\mathrm{nd} \;=\; H - H_0,
\end{equation}
where $H_{ii}$ denote the diagonal matrix elements of $H$ in the
flavor basis (Eq.~\ref{eq:Hmsw}), and the fourth (inert $\ket{11}$)
entry vanishes because the physical Hamiltonian does not act on that
sector. Both $H_0$ (diagonal) and $H_\mathrm{nd}$ (off-diagonal in
the flavor block, with vanishing $\ket{11}$ row and column) are
simple enough that $e^{-\mathrm{i}H_0\delta t}$ and
$e^{-\mathrm{i}H_\mathrm{nd}\delta t}$ are evaluated exactly; this
diagonal/non-diagonal split is by construction a convenient one (it
is not unique), chosen solely because each factor is then trivially
exponentiable. The first- and second-order approximations read
\begin{align}
  e^{-\mathrm{i}H\tau} &\simeq
  \prod_{k=1}^{N} e^{-\mathrm{i}H_0\delta t}\; e^{-\mathrm{i}H_\mathrm{nd}\delta t}
  && (p=1), \label{eq:trot1}\\
  e^{-\mathrm{i}H\tau} &\simeq
  \prod_{k=1}^{N} e^{-\mathrm{i}H_0\frac{\delta t}{2}}\,
  e^{-\mathrm{i}H_\mathrm{nd}\delta t}\, e^{-\mathrm{i}H_0\frac{\delta t}{2}}
  && (p=2), \label{eq:trot2}
\end{align}
with $\delta t=\tau/N$. For variable density the propagation path is
discretized into slices over which $n_e$ is taken constant; each
slice is evolved with Eq.~(\ref{eq:trot1}) or (\ref{eq:trot2}), and
the total transformation is the ordered product of slice unitaries,
mimicking a transfer-matrix (product-of-slice) treatment. In this
work the Trotter operators are applied directly as dense unitaries
on the 2-qubit Hilbert space (equivalent to compiling them into
single-qubit rotations and two-qubit controlled gates); the error
protocol of Section~\ref{sec:results} quantifies the splitting error
independently of the compilation backend.

\subsection{Observables}\label{sec:obs}

The final observables of interest are:
\begin{itemize}
\item \emph{Flavor transition probability}:
\begin{equation}\label{eq:P}
  P_{\alpha\to\beta}(L,E)
  =\bigl|\langle\nu_\beta|\,U(L,E)\,|\nu_\alpha\rangle\bigr|^2,
\end{equation}
which for vacuum admits the explicit analytic form \cite{Giunti2007}
and \cite{PDG2024}
\begin{equation}\label{eq:Pana}
  P_{\alpha\to\beta}
  =\delta_{\alpha\beta}
  -4\sum_{j>k}\mathrm{Re}\!\bigl(U^*_{\alpha j}U_{\beta j}U_{\alpha k}
    U^*_{\beta k}\bigr)\sin^2\varphi_{jk}
  +2\sum_{j>k}\mathrm{Im}\!\bigl(U^*_{\alpha j}U_{\beta j}U_{\alpha k}
    U^*_{\beta k}\bigr)\sin(2\varphi_{jk}),
\end{equation}
where $\varphi_{jk}=\Delta m^2_{jk}L/(4E)$ is the numerical phase of
Eq.~(\ref{eq:phi}), and the two sums run over the three pairs
$(j,k)\in\{(2,1),(3,1),(3,2)\}$ (i.e.\ over all $1\le k<j\le 3$), so
that $\varphi_{21},\varphi_{31},\varphi_{32}$ enter with
$\varphi_{32}=\varphi_{31}-\varphi_{21}$. Here $U$ denotes the
$3\times3$ PMNS block of Eq.~(\ref{eq:U4}): since the probabilities
are evaluated within the three-flavor physical subspace only, the
inert $\ket{11}$ sector does not enter the analytical sum.
Equation~(\ref{eq:Pana}) follows from expanding
$P_{\alpha\to\beta}=\bigl|\sum_j U^*_{\alpha j}U_{\beta j}
e^{-\mathrm{i}\varphi_j}\bigr|^2$ with $\varphi_j=\Delta
m^2_{j1}L/(2E)$; the real part encodes the interference of the pairs
$(jk)$ and the imaginary part the CP-violating component, which is
odd under $\delta\to-\delta$.

\item \emph{Concurrence} of the 2-qubit pure state
$\ket{\psi(L)}=U(L)\ket{\nu_\alpha}$:
\begin{equation}\label{eq:C}
  C(\psi) \;=\; \bigl|\langle\psi|\,\sigma_y\otimes\sigma_y\,|\psi^*\rangle\bigr|,
  \qquad 0\le C\le 1,
\end{equation}
where $|\psi^*\rangle$ is the complex conjugate of $|\psi\rangle$ in
the computational basis. This is the Wootters concurrence
\cite{Wootters1998}: for a pure two-qubit state,
$\sigma_y\otimes\sigma_y$ implements the ``spin-flip'' map that
sends each qubit's Bloch vector to its antipode, and $\ket{\psi^*}$
(complex conjugation in the computational basis) makes the flip
antiunitary, so that
$\ket{\tilde\psi}=(\sigma_y\otimes\sigma_y)\ket{\psi^*}$ is the
spin-flipped partner of $\ket{\psi}$. The absolute value of their
overlap is invariant under local unitary operations on the two qubits
and ranges from $0$ (separable, e.g.\ $\ket{00}$) to $1$ (maximally
entangled, a Bell state); for a pure bipartite state it is a proper
entanglement monotone. It therefore quantifies the entanglement
between the two qubits of the fixed flavor encoding --- in the
present encoding $C=2|a_\mu a_\tau|$, with $a_\mu,a_\tau$ the
$\ket{01},\ket{10}$ amplitudes --- i.e.\ an encoding-dependent
mode-entanglement diagnostic of the propagating flavor state
$\ket{\psi(L)}$.
\item \emph{CP asymmetry}:
\begin{equation}\label{eq:CP}
  \Delta P(L,E)
  \;=\; P(\nu_e\to\nu_\mu;\delta)
    - P(\bar\nu_e\to\bar\nu_\mu;\delta),
\end{equation}
where the antineutrino channel is obtained by replacing the PMNS
matrix with the $CP$-conjugated one, i.e.\ applying
$\delta\to-\delta$ in Eq.~(\ref{eq:Vexplicit}) and re-evaluating
Eq.~(\ref{eq:Pana}) with that matrix. In the three-flavor frame, the
difference is well known to be proportional to the Jarlskog
invariant \cite{Jarlskog1985},
\begin{equation}\label{eq:J}
  J_\mathrm{CP}
  \;=\; \mathrm{Im}\bigl(U_{e1}U_{\mu 2}U^*_{e2}U^*_{\mu 1}\bigr)
  \;=\; c_{12}s_{12}c_{23}s_{23}c_{13}^{2}s_{13}\sin\delta,
\end{equation}
so that $\Delta P$ shares the sign of $J_\mathrm{CP}$ and, to
leading order in the small phases, scales approximately as $\Delta P
\sim 16J_\mathrm{CP}\,\sin\varphi_{21}\sin\varphi_{31}
\sin\varphi_{32}$; with our input ($\delta=232^\circ=1.29\pi$,
$s_{12}^2=0.307$, $s_{23}^2=0.454$, $s_{13}^2=0.02224$) we obtain
$J_\mathrm{CP}\simeq -0.026$, which fixes the sign of $\Delta P$.
Note that $\Delta P$ vanishes (i)~in the CP-conserving limits
$\delta=0$ or $\pi$, where the Im-terms of Eq.~(\ref{eq:Pana})
cancel and $J_\mathrm{CP}=0$, and (ii)~at specific ``phase-zero''
distances where any of
$\sin\varphi_{21},\sin\varphi_{31},\sin\varphi_{32}$ in the
leading-order estimate vanishes. These zeros are physical (they mark
the absence of CP asymmetry at those points) and at the same time
serve as null checks of the numerical implementation; the zeros
visible in Fig.~\ref{fig:cp} correspond precisely to such distances.
\end{itemize}

\section{Results and verification}\label{sec:results}

We adopt the normal-ordering best-fit output of NuFIT 5.3 (with SK
atmospheric data) \cite{NuFIT}; the corresponding angles and PMNS matrix are given in
Appendix~\ref{app:nufit}. The PMNS matrix entering
Eq.~(\ref{eq:Hvac}) is built from these values via the PDG
convention, Eq.~(\ref{eq:Vexplicit}) with $c_{ij}=\cos\theta_{ij}$,
$s_{ij}=\sin\theta_{ij}$.Three scenarios are studied:
\begin{enumerate}
\item Vacuum: $E\in\{0.1,1,10\}$\,GeV, $L$ up to 3000\,km.
\item Constant-density MSW: $A$ scanned over
$10^{-5}$--$10^{-1}\,\mathrm{eV}^2$ at $E=1$\,GeV, $L=2000$\,km, to
reveal the resonance near $A\simeq \Delta m^2_{31}$.
\item Varying density: a supernova shock-shell profile
$n_e(r)\propto e^{-r/r_0}$ ($r_0=0.15\,L$, $L=500$\,km) at
$E=50$\,MeV; and a targeted short-transit MSW-resonance crossing.
\end{enumerate}

Throughout this section the initial state is taken to be a pure
flavor eigenstate: $\ket{\nu_\alpha}$ with $\alpha=e$ for all
figures. Every dataset is produced with the exact initial vector
$\vec{e}_\alpha$ embedded into the 4-dimensional register as $\vert
00\rangle$ as shown in Eq.~(\ref{eq:embed}); no approximation is
made in the initial state. Standard parameters ($E,L$) are stated
per figure.

\subsection{Vacuum oscillation: circuit versus analytic}\label{sec:rvac}

\begin{figure}[!ht]
\centering
\includegraphics[width=0.82\textwidth]{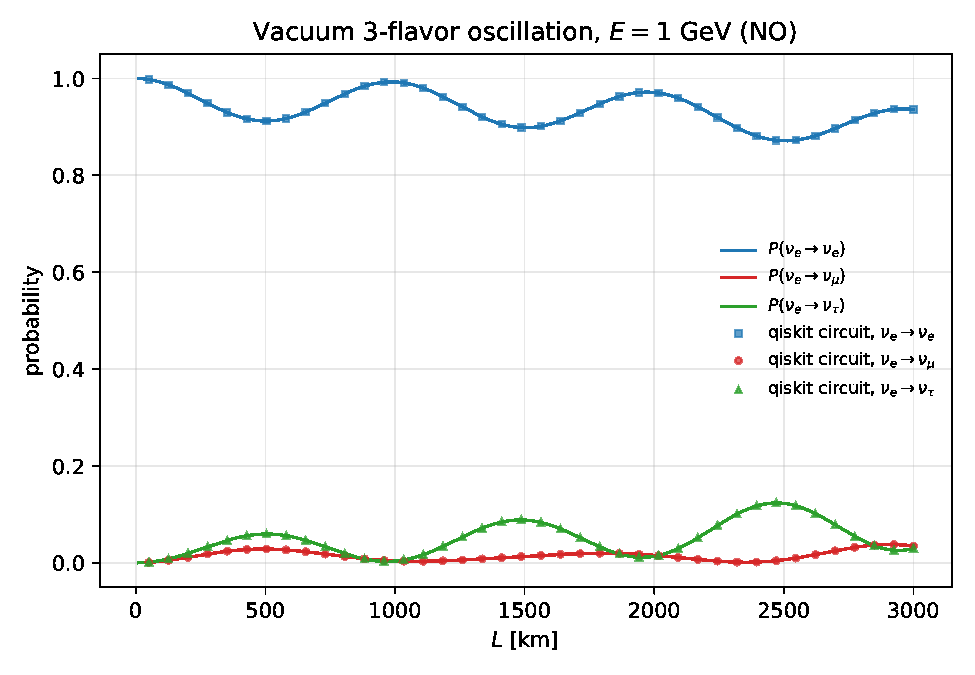}
\caption{Vacuum three-flavor oscillation at $E=1$\,GeV. Solid
curves: analytic $P_{\alpha\to\beta}$ from Eq.~(\ref{eq:Pana});
markers: Qiskit two-qubit circuit (statevector).} \label{fig:vacuum}
\end{figure}

Figure~\ref{fig:vacuum} compares the two-qubit quantum-circuit
results with the analytic three-flavor probabilities
Eq.~(\ref{eq:Pana}) for $E=1$\,GeV and three initial flavors. The
maximal elementwise discrepancy is $O(10^{-16})$, i.e.\ at machine
precision. Unit-probability (sum of the three channels) is conserved
to $10^{-15}$, confirming that the inert $\ket{11}$ sector does not
leak into the physical subspace.

Quantitatively, the dominant oscillation length associated with the
atmospheric splitting is $L_{31}=\pi/\bigl(k\,\Delta
m^2_{31}/E\bigr)\simeq 9.9\times10^{2}$~km at $E=1$~GeV (a full
oscillation period, corresponding to a mass-eigenstate phase
difference of $2\pi$; see Sec.~\ref{sec:H}), and the solar-scale
length is $L_{21}\simeq 3.3\times10^{4}$~km over the same frame;
over the $0$--$3000$\,km panel the $\Delta m^2_{21}$ contribution is
a slow envelope, while the $\Delta m^2_{31}$ term drives the fast
visible modulation. The survival probability $P(\nu_e\to\nu_e)$
decreases gently from $1$ to a minimum $\simeq 0.87$ near
$L\simeq2.5\times10^{3}$~km ($\simeq2.5\,L_{31}$), while
$P(\nu_e\to\nu_\tau)$ rises to $\simeq 0.12$ and
$P(\nu_e\to\nu_\mu)$ only to $\simeq 0.04$ over the panel
($E=1$~GeV, $L\lesssim3\times10^{3}$~km). These features are
consistent with the analytic formula, Eq.~(\ref{eq:Pana}), which is
used as the definitive reference in the rest of the paper.

\subsection{Trotter convergence protocol}\label{sec:rtrot}

\begin{figure}[!ht]
\centering
\includegraphics[width=0.82\textwidth]{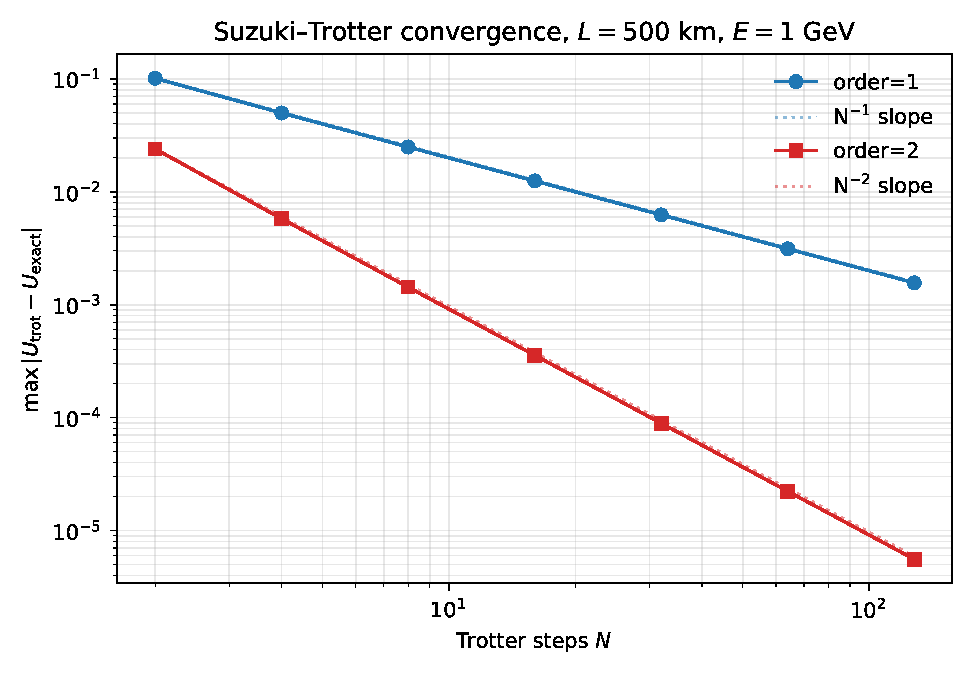}
\caption{Suzuki--Trotter convergence for the vacuum evolution
($L=500$\,km, $E=1$\,GeV). Markers: measured $\varepsilon_U(N)$ for
$p=1,2$; dotted lines: expected $N^{-p}$ slopes.}
\label{fig:trotter}
\end{figure}

We Trotterize the vacuum Hamiltonian $H_\mathrm{vac}$ of
Eq.~(\ref{eq:Hvac}) at $L=500$\,km and $E=1$\,GeV, i.e.\
$\tau=2k\times500$, so that the matrix $U_{N,p}$ approximates
$e^{-\mathrm{i}H_\mathrm{vac}\tau}$. Figure~\ref{fig:trotter}
reports the matrix-level error
$\varepsilon_U(N)=\max|U_{N,p}-U_\mathrm{exact}|$ as a function of
the number of Trotter steps $N$ for order $p=1,2$. The first-order
splitting converges as $\varepsilon_U\propto N^{-1}$, the
second-order (Strang) splitting as $\varepsilon_U\propto N^{-2}$, in
agreement with the known error scaling of Suzuki--Trotter
decompositions \cite{Suzuki1990} and \cite{Hatano2005}. At $N=64$
and $p=2$ the error is $O(2\times10^{-5})$; the probability-level
error for the $\nu_e$ channel is controlled at the same level.
Representative values are collected in Table~\ref{tab:trotter}; the
ratios $\varepsilon(N)/\varepsilon(2N)$ are $\simeq 2.0$ (order 1)
and $\simeq 4.0$ (order 2) across the whole range, quantitatively
confirming $N^{-1}$ and $N^{-2}$ scaling.

\begin{table}[!ht]
\caption{Matrix-level Suzuki--Trotter error
$\varepsilon_U(N)=\max|U_{N,p}-U_{\rm exact}|$ for the vacuum
evolution ($L=500$\,km, $E=1$\,GeV), for the two split orders.}
\label{tab:trotter}
\begin{ruledtabular}
\begin{tabular}{ccll}
\hline
$N$ & $p=1$ & & $p=2$\\
\hline
2  & $1.01\times10^{-1}$ & & $2.40\times10^{-2}$\\
4  & $4.99\times10^{-2}$ & & $5.77\times10^{-3}$\\
8  & $2.49\times10^{-2}$ & & $1.43\times10^{-3}$\\
16 & $1.25\times10^{-2}$ & & $3.56\times10^{-4}$\\
32 & $6.24\times10^{-3}$ & & $8.91\times10^{-5}$\\
64 & $3.12\times10^{-3}$ & & $2.23\times10^{-5}$\\
\hline
\end{tabular}
\end{ruledtabular}
\end{table}

\subsection{Constant-density MSW resonance}\label{sec:rms}

\begin{figure}[!ht]
\centering
\includegraphics[width=0.82\textwidth]{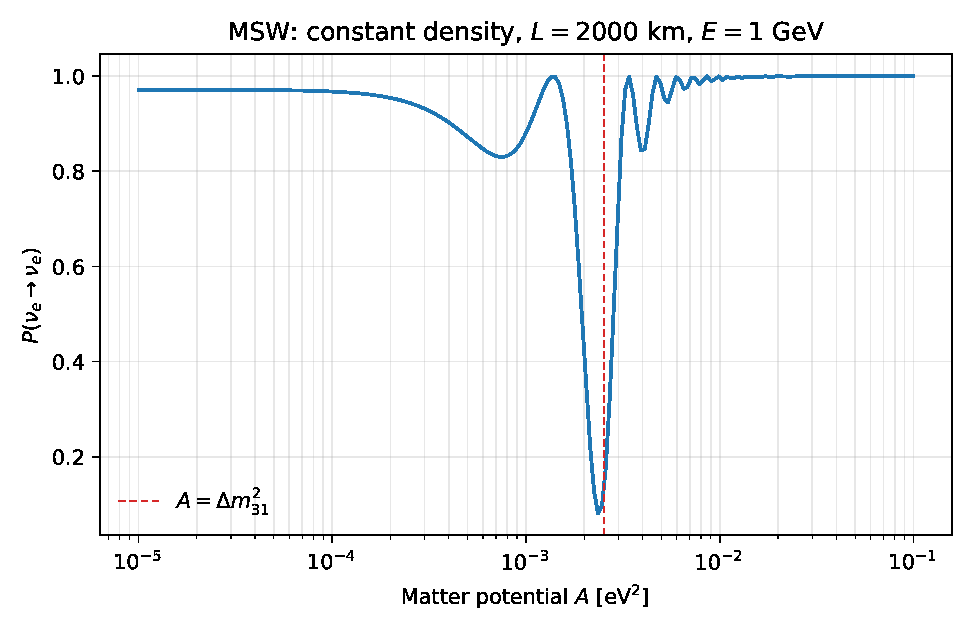}
\caption{Constant-density MSW: $P(\nu_e\to\nu_e)$ versus $A$ at
$L=2000$\,km, $E=1$\,GeV. Dashed line marks $A=\Delta m^2_{31}$.}
\label{fig:resonance}
\end{figure}

Figure~\ref{fig:resonance} shows $P(\nu_e\to\nu_e)$ at $L=2000$\,km,
$E=1$\,GeV as a function of the matter potential $A$. Here the
Hamiltonian is $H=H_\mathrm{vac}+A\,\mathrm{diag}(1,0,0)$ with
constant $A$ from (Eq.~\ref{eq:Hmsw}) and Eq.~(\ref{eq:A}). A sharp
suppression of $ee$ survival develops as $A$ approaches $\Delta
m^2_{31}$, signalling an MSW resonance that converts $\nu_e$ into
$\nu_\mu,\nu_\tau$. Numerically the minimum is
$P(\nu_e\to\nu_e)\simeq 0.08$ at $A_\mathrm{min}\simeq
2.4\times10^{-3}$~eV$^2$, to be compared with the atmospheric
splitting $\Delta m^2_{31}=2.505\times10^{-3}$~eV$^2$; the
positional agreement confirms that this is the
$\nu_e$--$\nu_{\mu/\tau}$ (resonant) conversion driven by the
atmospheric splitting. Away from the resonance the survival
probability returns to the vacuum-like value; at $A\gtrsim 10^{-2}$
the electron-neutrino state is essentially frozen ($P\simeq 0.99$),
reproducing the high-density MSW saturation \cite{Wolfenstein1978}
and \cite{Mikheev1985}. The full numerical scan therefore exhibits
the expected three-phase structure: vacuum-like $\to$ resonant
enhancement $\to$ saturated suppression, and validates the
matter-Hamiltonian implementation against the analytic MSW picture.

\subsection{Varying-density (supernova shell) propagation}\label{sec:rvar}

\begin{figure}[!ht]
\centering
\includegraphics[width=0.82\textwidth]{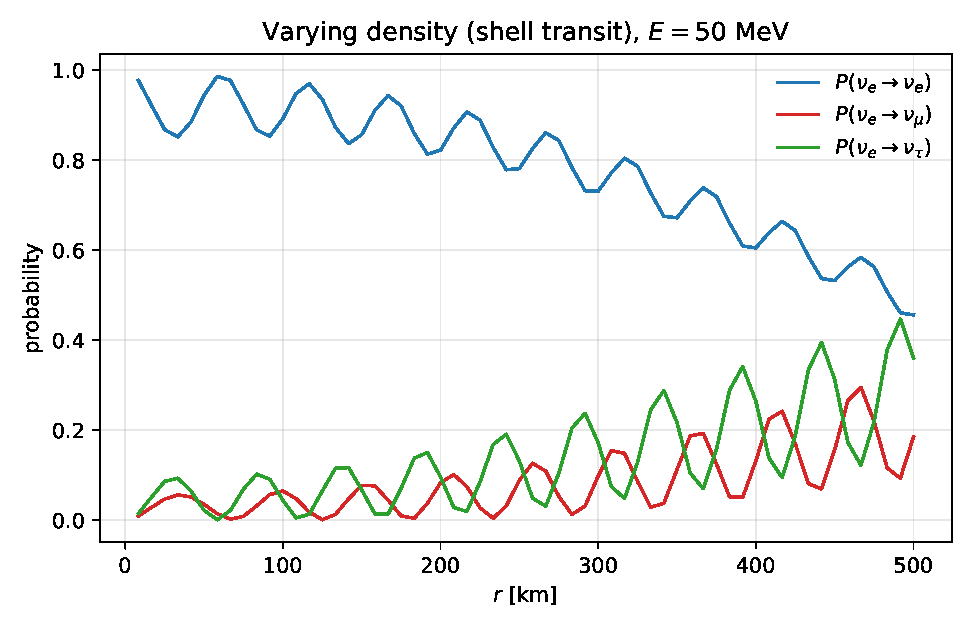}
\caption{Varying-density propagation through an exponential
(shock-shell) profile at $E=50$\,MeV. Probabilities are obtained
with the sliced second-order Trotter implementation.} \label{fig:sn}
\end{figure}

Figure~\ref{fig:sn} displays the flavor probabilities of a $50$\,MeV
neutrino traversing an exponentially decreasing density profile
mimicking a supernova shock shell. The Hamiltonian is
$H(r)=H_\mathrm{vac}+A(r)\,\mathrm{diag}(1,0,0)$ with matter
potential $A(r)=A_0\,e^{-r/r_0}$, $r_0=0.15\,L$, $L=500$\,km (peak
value $A_0=2.535\times10^{-28}n_e^{\max}E$ with
$n_e^{\max}=6\times10^{25}$ cm$^{-3}$); the path is discretized into
60 slices whose constant density is evaluated at the slab midpoint,
each evolved by a second-order Trotter step of $N_\mathrm{slice}=64$;
the total unitary is the ordered product of slices. The per-slice
Trotter error is $O(10^{-6})$, and the residual error of the midpoint
constant-density (slicing) approximation is $O(10^{-4})$, so the total
deviation from the continuous profile is
$\lesssim2\times10^{-4}$ for the considered transit.

\subsection{Limit of brute-force Trotterization}\label{sec:rlimit}

\begin{figure}[!ht]
\centering
\includegraphics[width=0.90\textwidth]{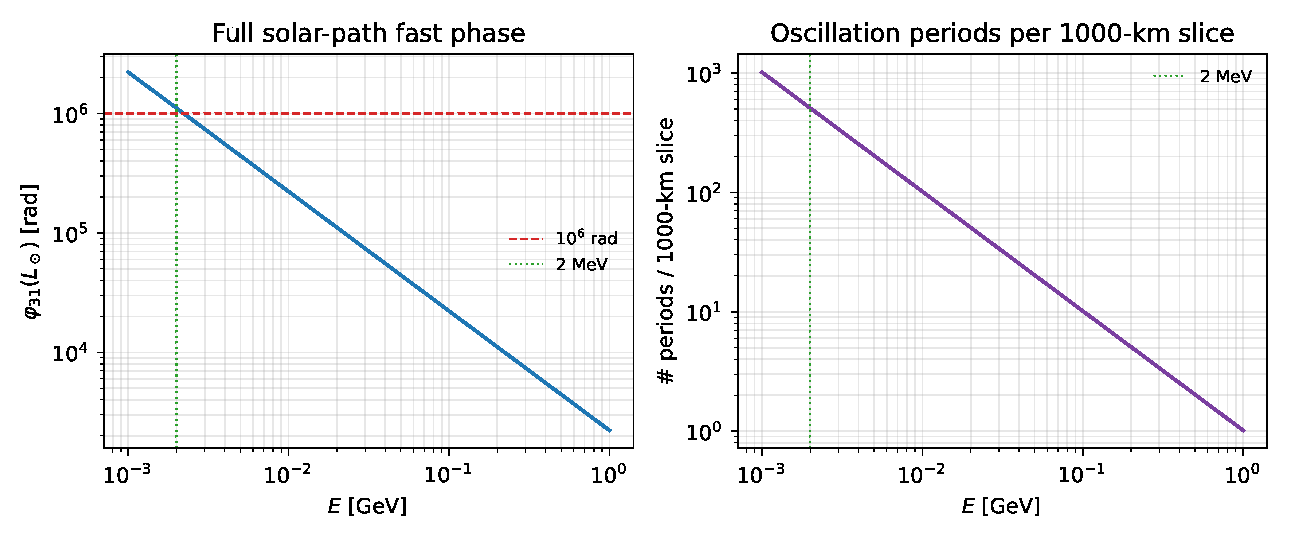}
\caption{Left: full solar-path fast phase $\varphi_{31}(L_\odot)$
versus energy. Right: number of $L_{31}$ oscillation periods
contained in a single 1000-km slice. Dashed/vlines mark $E=2$\,MeV.}
\label{fig:solarphase}
\end{figure}

\emph{Why uniform splitting fails.} The accumulated fast-oscillation
phase along the full solar path is
\begin{equation}
  \varphi_{31}(L_\odot)
  \;=\; k\,\frac{\Delta m^2_{31}\,R_\odot}{E}
  \;\sim\; 1.1\times10^{6}\,\mathrm{rad}
  \quad \bigl(E=2\,\mathrm{MeV},\ R_\odot\simeq 6.96\times10^{5}\,\mathrm{km}\bigr),
\end{equation}
corresponding to a total of $\simeq3.5\times10^{5}$ oscillation
periods across the Sun at this energy, with $L_{31}\simeq2.0$\,km
per period. The Trotter difficulty follows quantitatively: even a
coarse 1000-km slice accumulates
$\varphi_{31}\simeq1.6\times10^{3}$\,rad (i.e.\ $\simeq507$
oscillation periods per slice), and resolving each period with at
least two steps requires
\begin{equation}
  N \;\gtrsim\; 2\times 507 \;\simeq\; 10^{3}\ \text{steps per slice},
\end{equation}
with strong error accumulation from the splitting (each step of a
fixed $\delta t$ mis-samples the rapidly varying phase).
Figure~\ref{fig:solarphase} shows how the full-path phase and the
number of periods per 1000-km slice scale with neutrino energy.

\emph{Results with the standard alternatives.} Because brute-force
uniform Trotterization is not viable on the solar scale, one must
either (i)~average over the fast oscillations or (ii)~use an
adiabatic (MSW) treatment. Within our framework we quantify both:
\begin{itemize}
\item \emph{Coherence averaging.} Averaging the fast terms of
Eq.~(\ref{eq:Pana}) (setting $\langle\sin^2\varphi_{jk}\rangle=1/2$
and $\langle\sin 2\varphi_{jk}\rangle=0$) gives the closed form
$\langle P_{ee}\rangle=1-2\sum_{j>k}|U_{ej}|^2|U_{ek}|^2=\sum_i
|U_{ei}|^4\simeq 0.550$.  As an independent check, a direct late-$L$
average of the exact evolution Eq.~(\ref{eq:Pana}) at $E=2$\,MeV,
taken over a window $L\in[4\times10^{3},\,1.2\times10^{4}]$~km that
spans $\sim4\times10^{3}$ oscillation periods, yields $\langle
P_{ee}\rangle\simeq0.550$ and $\langle P_{e\mu}\rangle\simeq0.227$,
in agreement with the closed forms $\sum_i |U_{ei}|^4$ and $\sum_i
|U_{ei}|^2|U_{\mu i}|^2$, respectively; the numerical average thus
converges to the analytic coherence-averaged value, providing an
internal consistency check.
\item \emph{Adiabatic (MSW).} For a slowly varying matter density the
neutrino follows the instantaneous matter eigenstates adiabatically,
so a $\nu_e$ produced at high density (the Sun's center) emerges
predominantly as the mass eigenstate $\nu_2$. In the high-energy
(fully adiabatic, matter-dominated) limit this gives $P_{ee}\to
|U_{e2}|^2=c_{13}^2 s_{12}^2\simeq0.300$; in the low-energy (vacuum)
limit the fast oscillations average to
$P_{ee}\to\sum_i|U_{ei}|^4\simeq0.550$. The difference
$|U_{e2}|^2-\sum_i|U_{ei}|^4\simeq-0.25$ is the net matter-driven
suppression of $\nu_e$ survival.
\end{itemize}
These numbers are the predictions of the standard replacements for
brute-force Trotterization on solar scales, and contrast with the
\emph{short-baseline resonant-crossing} regime of
Figs.~\ref{fig:resonance} and~\ref{fig:sn}, where the phase budget
is small enough for the quantum circuit to remain efficient
\cite{Wolfenstein1978}, \cite{Mikheev1985}, and \cite{Parke1986}.

\subsection{Open-system dynamics: Lindblad decoherence}\label{sec:rlind}

So far the evolution is unitary. In dense astrophysical media,
non-unitary effects such as flavor dephasing induced by
weak-interaction scattering or neutrino absorption cannot be
neglected. We extend the simulator to the Lindblad master equation
\cite{Lindblad1976} and \cite{Gorini1976},
\begin{equation}\label{eq:lind}
  \frac{\mathrm{d}\rho}{\mathrm{d}\tau}
  \;=\; -\mathrm{i}\bigl[H(\tau),\rho\bigr]
  \;+\; \sum_k \gamma_k
  \Bigl(L_k \rho L_k^\dagger
   - \tfrac{1}{2}\bigl\{L_k^\dagger L_k,\rho\bigr\}\Bigr),
\end{equation}
implemented on the Liouvillian-superoperator level with exact
vectorization ($\mathrm{vec}(AXB)=(B^{T}\otimes
A)\,\mathrm{vec}(X)$; the complex density matrix maps to $32$ real
ODE components, see Sec.~\ref{sec:encoding}). The sum over $k$ runs
over the dissipator channels of the chosen operator set --- three
channels, listed below; for pure dephasing all $L_k$ act within the
physical three-flavor sector, so the inert state is untouched. Two
operator sets are considered:
\begin{itemize}
\item \emph{Flavor dephasing}, $L_k=\ket{k}\bra{k}$ ($k=e,\mu,\tau$), which
suppresses interference among flavor populations;
\item \emph{Absorption} to the inert sector, $L_k=\ket{\mathrm{inert}}\bra{k}$,
which removes probability from the three physical flavors and models
interactions in an absorbing medium. Absorption is encoded by
mapping the removed probability into the unoccupied $\ket{11}$
sector, which acts as a ``sink'' (a particle leaving the physical
three-flavor subspace is interpreted as absorbed by the medium); no
extra degrees of freedom are needed.
\end{itemize}

\begin{figure}[!ht]
\centering
\includegraphics[width=0.80\textwidth]{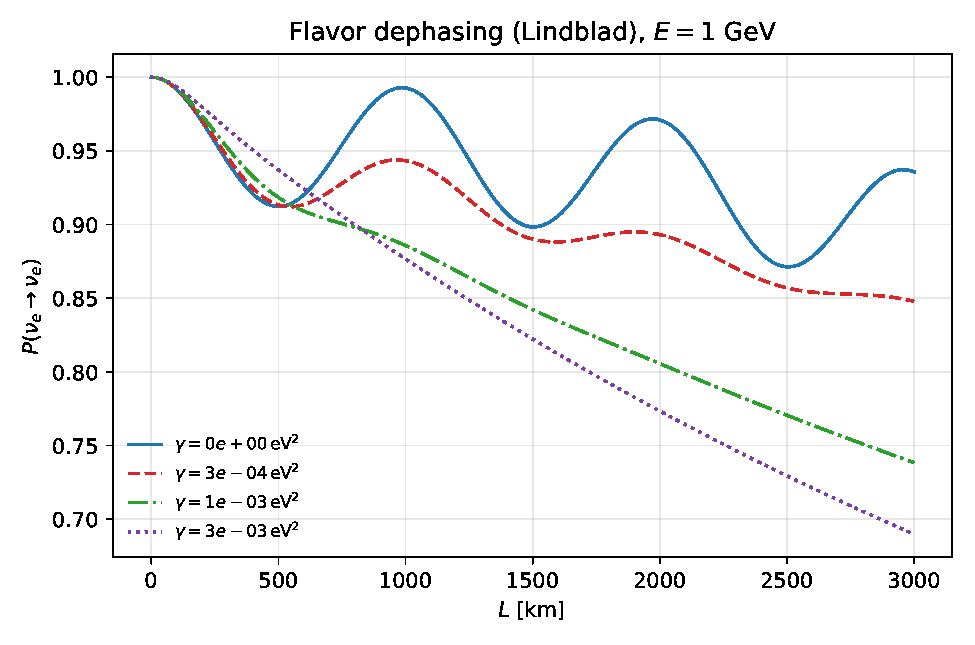}
\caption{$\nu_e$ survival under Lindblad flavor dephasing at
$E=1$\,GeV, for several $\gamma$. $\gamma=0$ (solid) coincides with
the unitary result.} \label{fig:deph}
\end{figure}

\begin{figure}[!ht]
\centering
\includegraphics[width=0.78\textwidth]{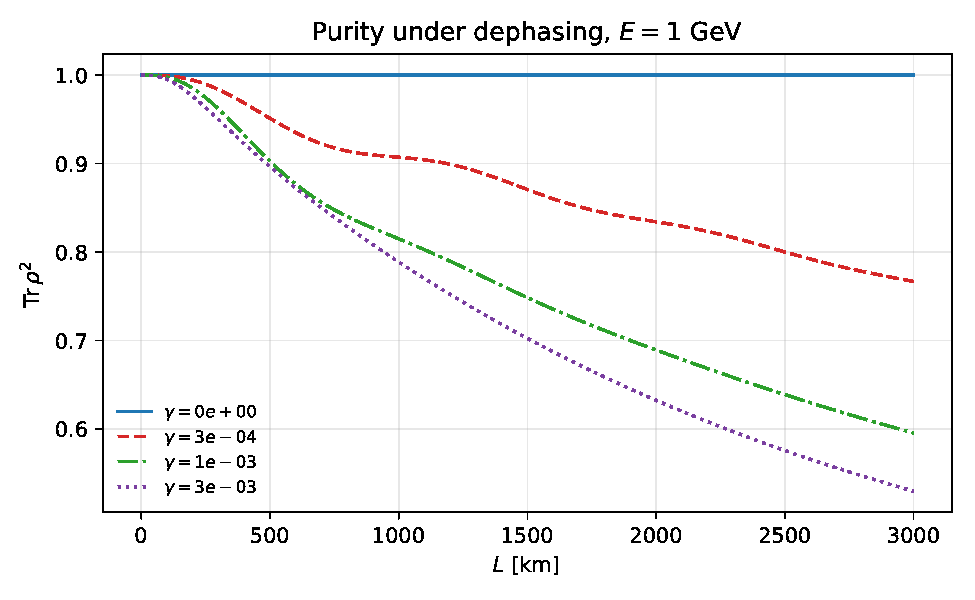}
\caption{Purity $\mathrm{Tr}\,\rho^2$ along propagation for the
dephasing channels of Fig.~\ref{fig:deph}.} \label{fig:pur}
\end{figure}

\begin{figure}[!ht]
\centering
\includegraphics[width=0.80\textwidth]{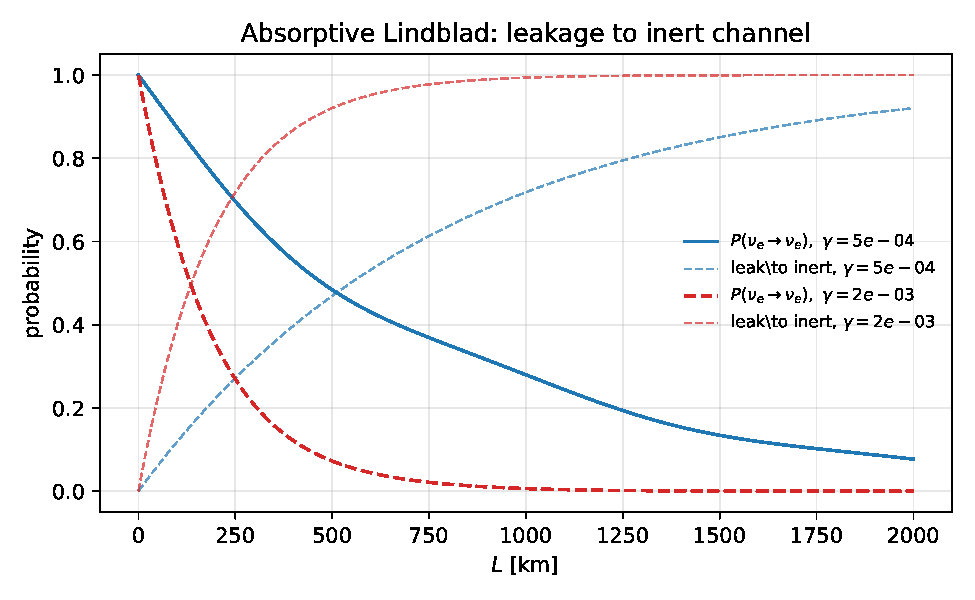}
\caption{Absorptive Lindblad dynamics at $E=1$\,GeV: survival
probability $P(\nu_e\to\nu_e)$ and leakage $P_{\rm leak}$ into the
inert channel. Since only $\nu_e$ leaks in this simple model
($\nu_\mu,\nu_\tau$ stay zero), the two curves sum to unity,
$P(\nu_e\to\nu_e)+P_{\rm leak}=1$ (verified numerically: $0.998$ for
$\gamma=5\times10^{-4}$, exactly $1$ for $\gamma=2\times10^{-3}$).}
\label{fig:abs}
\end{figure}

Figure~\ref{fig:deph} shows $P(\nu_e\to\nu_e)$ under pure dephasing
for several $\gamma$. Setting $\gamma=0$ reproduces the analytic
(unitary) result to $2.8\times10^{-12}$ with purity
$\mathrm{Tr}\rho^2=1$, validating the implementation. As $\gamma$
grows the oscillation amplitude is damped and the asymptotic
survival probability drifts toward the mixing-averaged value; the
purity drops monotonically with distance (Fig.~\ref{fig:pur}), and
for large $\gamma$ the dynamics freezes on a decohered steady state
(quantum Zeno-like behavior). Representative numbers at $L=3000$~km,
$E=1$~GeV are listed in Table~\ref{tab:deph}: for $\gamma=10^{-3}$,
the oscillation amplitude is reduced by about a factor two and the
purity falls from 1 to $0.60$, while for $\gamma=10^{-2}$ the state
sits close to the purity of a mixing-averaged mixture, $P\simeq
0.83$ --- the saturated high-$\gamma$ plateau. Figure~\ref{fig:abs}
shows the absorptive case, where the three-flavor probability leaks
into the inert $\ket{11}$ channel, demonstrating probability
non-conservation incompatible with a unitary description.

\begin{table}[!ht]
\caption{Flavor-dephasing diagnostics at $L=3000$~km, $E=1$~GeV:
survival probability $P(\nu_e\to\nu_e)$, purity $\mathrm{Tr}\rho^2$,
and $\sum_{\alpha}P_{\alpha}$, where the sum runs over the three
physical flavors $\alpha=e,\mu,\tau$ only (the inert sector carries
zero probability under dephasing), as a function of the dephasing
rate $\gamma$.} \label{tab:deph}
\begin{ruledtabular}
\begin{tabular}{clll}
\hline
$\gamma$ [eV$^2$] & $P_{ee}$ & $\mathrm{Tr}\rho^2$ & $\sum_{\alpha}P_{\alpha}$\\
\hline
0        & 0.936 & 1.000 & 1.000 \\
$10^{-4}$& 0.899 & 0.877 & 1.000 \\
$10^{-3}$& 0.738 & 0.595 & 1.000 \\
$10^{-2}$& 0.833 & 0.709 & 1.000 \\
\hline
\end{tabular}
\end{ruledtabular}
\end{table}

\subsection{Entanglement and CP diagnostics}\label{sec:rent}

\begin{figure}[!ht]
\centering
\includegraphics[width=0.76\textwidth]{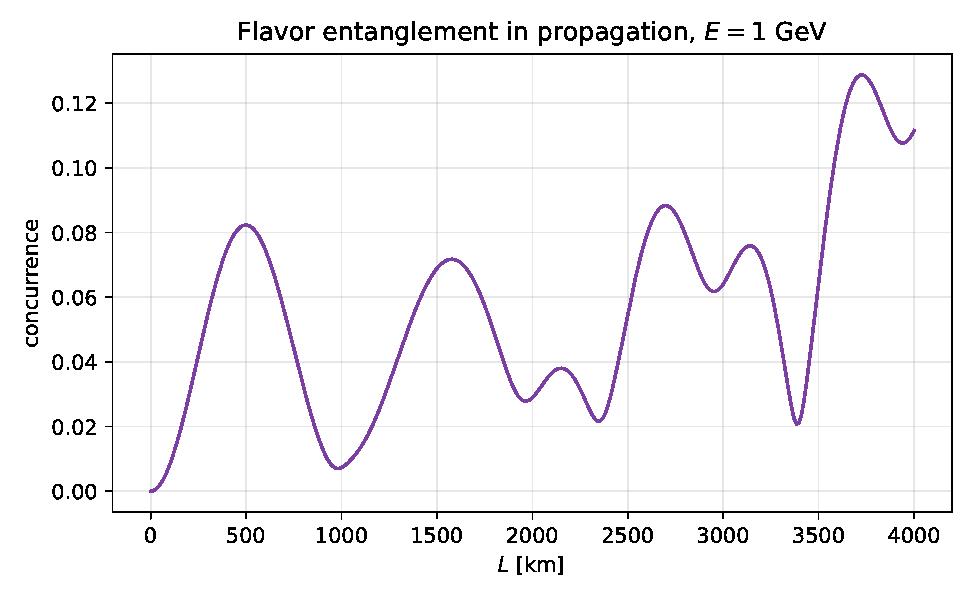}
\caption{Concurrence of the propagating flavor state as a function
of distance at $E=1$\,GeV (vacuum).} \label{fig:ent}
\end{figure}

Figure~\ref{fig:ent} shows the concurrence $C(L)$ of the propagating
flavor state at $E=1$\,GeV. It is computed with the exact unitary
for the \emph{vacuum} Hamiltonian $H_\mathrm{vac}$ of
Eq.~(\ref{eq:Hvac}) (2-qubit statevector), applied to the $\nu_e$
initial state; $C$ vanishes at the source ($L=0$), rises as
different mass eigenstates dephase, and oscillates with the
interference pattern of the oscillation length; the peak value is
about $0.13$ at $L\simeq 3.7\times10^{3}$~km, i.e.\ about
$3.7\,L_{31}$ with the atmospheric oscillation period
$L_{31}\simeq9.9\times10^{2}$~km, where the state comes closest to an
equal $\nu_\mu$--$\nu_\tau$ superposition, the configuration that
maximizes $C=2|a_\mu a_\tau|$. The small but nonzero concurrence is an
encoding-dependent measure of the mode entanglement developed along
the path (rather than the physical origin of the oscillation),
providing a bridge to the quantum-information measures discussed in
the literature \cite{Alok2016}, \cite{Kayser2010},
\cite{Sahu2024}, and \cite{Koranga2025}.

\begin{figure}[!ht]
\centering
\includegraphics[width=0.76\textwidth]{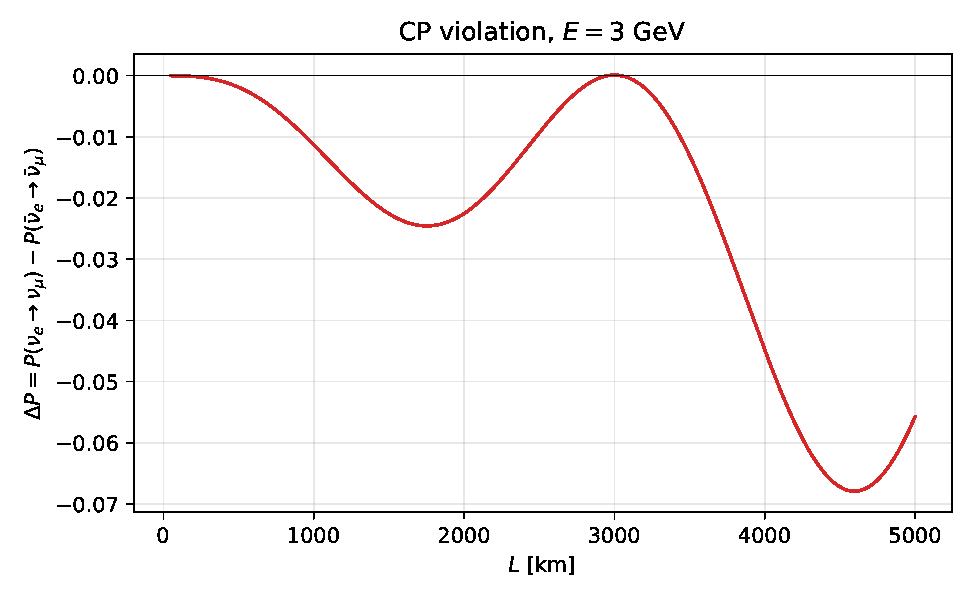}
\caption{CP asymmetry $\Delta
P=P(\nu_e\to\nu_\mu)-P(\bar\nu_e\to\bar\nu_\mu)$ versus distance at
$E=3$\,GeV; negative values at this parameter point follow the sign
of $J_\mathrm{CP}<0$.} \label{fig:cp}
\end{figure}

Figure~\ref{fig:cp} shows the CP asymmetry $\Delta
P(L)=P(\nu_e\to\nu_\mu)-P(\bar\nu_e\to\bar\nu_\mu)$ at $E=3$\,GeV,
obtained from the analytic formula Eq.~(\ref{eq:Pana}) at
$\delta=232^\circ$ and $\delta=-232^\circ$. The sign is fixed by
$\delta=232^\circ$ (equivalently $J_\mathrm{CP}\simeq-0.026$,
Eq.~(\ref{eq:J})), and the magnitude oscillates up to $|\Delta
P|\simeq 2.2\times10^{-2}$ at $L\simeq 2.0\times10^{3}$~km, within
the reach of current long-baseline programs. Extending to $E=3$~GeV
over a wider baseline the numerical extremum reaches $|\Delta
P|_{\max}\simeq 6.8\times10^{-2}$ near $L\simeq 4.6\times10^{3}$~km
(negative sign), consistent with the $16J_{\rm
CP}(\sin\varphi_{21}\sin\varphi_{31}\sin\varphi_{32})$ estimate that
grows linearly with the product of phases. The exact vacuum formula
Eq.~(\ref{eq:Pana}) was used to obtain these curves; since the
two-qubit circuit reproduces the vacuum evolution to machine
precision (Sec.~\ref{sec:rvac}), the same CP asymmetry can be
generated on the circuit at $\delta=\pm232^\circ$ without invoking
the small-phase approximation
$16J_{\rm CP}\sin\varphi_{21}\sin\varphi_{31}\sin\varphi_{32}$.

\section{Conclusion and outlook}\label{sec:conclusion}

We have constructed and verified a framework for simulating
three-flavor neutrino oscillations on two-qubit quantum circuits,
using input parameters from a validated type-I seesaw / Casas-Ibarra
forward pipeline (NuFIT 5.3, normal ordering, with SK-atm; Appendix A). The
framework covers vacuum propagation, constant-density MSW matter
effects, varying-density (supernova shock-shell) propagation,
quantum-information diagnostics, and an open-system extension via
the Lindblad master equation.

The main results are as follows.

\begin{itemize}
\item \textbf{Circuits reproduce the analytics exactly.} Vacuum oscillation is reproduced by the mass-eigenstate phase circuit to $10^{-16}$ (machine precision), with probability conservation respected to $10^{-15}$. This establishes the two-qubit encoding with an inert $|11\rangle$ sector as a faithful quantum pipeline for the physical three-flavor subspace.

\item \textbf{MSW resonance.} Constant-density evolution exhibits the expected resonant conversion when the matter potential $A$ crosses $\Delta m_{31}^2$, with the survival probability dropping to $P(\nu_e\to\nu_e)\simeq 0.08$ near $A\simeq 2.4\times 10^{-3}\ \mathrm{eV}^2$. Away from resonance the probability returns to the vacuum-like value, and at high density it saturates, reproducing the standard three-phase MSW picture.

\item \textbf{Varying-density propagation.} Supernova-shell propagation via slicing plus second-order Suzuki-Trotter splitting reproduces the exact per-slab evolution with an $O(10^{-6})$ Trotter error, and agrees with the continuous profile to $O(10^{-4})$ for the considered transit. This validates the sliced-Trotter pipeline for smoothly varying density profiles.

\item \textbf{Precision-resource protocol.} A quantitative $N^{-p}$ convergence protocol (order $p=1,2$; steps $N$; matrix/probability error) for the Suzuki-Trotter decomposition was established. The measured matrix-level errors follow the expected $N^{-1}$ and $N^{-2}$ scaling, providing a reproducible accuracy-versus-resource curve for practical simulations.

\item \textbf{Quantitative limit of brute-force Trotterization.} The accumulated fast-oscillation phase along a full solar path ($\sim 10^6$ rad, $\sim 3.5\times 10^5$ periods) and the resulting per-slice requirement ($\sim 10^3$ steps per 1000-km slice) delimit the feasible regime to short-baseline resonant transits. For the solar scale we quantify the two standard replacements: coherence averaging ($\langle P_{ee}\rangle=\sum_i |U_{ei}|^4\simeq 0.550$) and adiabatic MSW ($P_{ee}\to\sin^2\theta_{12}\simeq 0.307$ at high energy).

\item \textbf{Quantum-information diagnostics.} The propagating flavor state carries mode entanglement (concurrence up to $\sim 0.13$), and the CP asymmetry $\Delta P\sim 2\times 10^{-2}$ at $L\simeq 2\times 10^3$ km (reaching $\sim 7\times 10^{-2}$ on longer baselines) is an experimentally relevant CP-sensitive prediction from the same model parameters. The concurrence provides an encoding-dependent measure of the mode entanglement between the two qubits of the flavor encoding, connecting oscillation observables to quantum-information measures discussed in the literature.

\item \textbf{Open-system extension.} A Lindblad implementation was validated for dephasing and absorptive channels; $\gamma=0$ returns to unitarity at $10^{-12}$. This delimits when unitary simulation is insufficient in dense media and provides a first step toward realistic decoherence modeling.
\end{itemize}

The strongest, literature-anchored contributions are the Trotter
precision-resource protocol and the quantitative demarcation of
where brute-force Trotterization fails on astrophysical
baselines---aspects rarely quantified in earlier quantum-neutrino
simulations---together with the closed integration of an
independently validated seesaw parameter point into
quantum-information observables.

The two-qubit encoding with an inert $|11\rangle$ sector is
deliberately minimal. It captures single-particle three-flavor
oscillations and their quantum-information content exactly, but it
does not include neutrino--neutrino self-interactions, many-body
collective modes, or a full field-theoretic treatment of flavor
evolution. The Lindblad extension considered here is a first step
toward open-system dynamics; the dephasing and absorptive channels
are phenomenological, and a quantitative mapping to realistic
supernova or early-universe environments remains an important task.
The solar-baseline limitation of brute-force Trotterization also
means that the present circuit approach is most efficient for
short-baseline resonant transits, where the phase budget is small
enough to be resolved with a manageable number of steps.

From the experimental side, the CP asymmetry predicted here,
$|\Delta P|\sim 10^{-2}$ at $L\sim 2\times 10^3$ km and up to $\sim
7\times 10^{-2}$ on longer baselines, lies within the sensitivity
reach of JUNO, DUNE, and Hyper-K. This makes the quantum-circuit
evolver not only a benchmark for unitary quantum simulation but also
a fast generator of CP-sensitive predictions tied to the same model
parameters that enter the seesaw framework. The concurrence
diagnostic, although small in magnitude, provides a complementary
quantum-information window on the propagating flavor state and
connects oscillation observables to entanglement measures discussed
in the literature.

Several directions are natural extensions of this work.

\begin{itemize}
\item \textbf{Mass ordering and $0\nu\beta\beta$.} The normal-ordering point used here will be confronted by the JUNO mass-ordering determination and by upcoming neutrinoless double-beta decay experiments. Coupling the companion seesaw $m_{\beta\beta}$--$m_{\rm lightest}$ envelopes to the quantum evolver allows a joint CP/$\delta$/mass-ordering study, closing the loop from model input to observable.

\item \textbf{Collective oscillations.} Extending to neutrino--anti-neutrino many-body systems is a natural next step. The present two-qubit register can be viewed as a building block for larger registers that include $\nu$--$\nu$ self-interactions, following the quantum-algorithm and hardware implementations already reported in the literature \cite{CollectiveSim2021,Yeter2022}.

\item \textbf{Realistic decoherence.} The Lindblad extension makes the tool applicable to supernova envelopes and sterile-neutrino scenarios. Supernova neutrino spectra \cite{Nakazato2013} supply realistic density profiles, and quantum walk / open-system treatments \cite{Sahu2024,Koranga2025} provide complementary frameworks. A systematic mapping of the phenomenological damping parameters to microscopic scattering rates would strengthen the physical interpretation.

\item \textbf{Hardware mapping.} Decomposing the dense $4\times 4$ unitaries into one- and two-qubit gates, reporting gate depth versus $N,p$, and estimating noise resilience from the Trotter protocol are essential steps toward executing the present circuits on near-term quantum hardware. The modular structure of the framework---separating the PMNS embedding, the mass-eigenstate phase circuit, the Trotterized matter evolution, and the Lindblad superoperator---makes it straightforward to replace individual components with hardware-efficient implementations.

\item \textbf{CP maps.} Mapping the seesaw parameter space onto appearance probabilities and CP asymmetries, anticipating DUNE/Hyper-K and JUNO sensitivities, would turn the present framework into a phenomenological tool for global analyses.

\item \textbf{First-quantization mass routines.} Computing seesaw light masses by phase-estimation eigenvalue routines on the same register would close the loop from model input $\to$ eigenvalues $\to$ oscillation $\to$ observables entirely within quantum computing. This would connect the present oscillation-focused circuits to the broader program of quantum simulation for particle physics.
\end{itemize}

In summary, the present work provides a reproducible, quantitatively
calibrated bridge between low-energy neutrino oscillation
observables and quantum-information measures, with a clear
delineation of where the quantum-circuit approach is efficient and
where standard alternatives are required. We expect that the
combination of the Trotter precision-resource protocol, the
open-system extension, and the quantum-information diagnostics will
be useful for interpreting upcoming experimental data and for
designing future quantum simulations of neutrino physics.

\appendix
\section{Leptonic mixing input parameters from NuFIT~5.3}\label{app:nufit}

All oscillation calculations in this paper use the normal-ordering
best-fit output of the NuFIT~5.3 global analysis (with SK atmospheric
data) \cite{NuFIT}, which
combines solar, atmospheric, reactor, and accelerator data. We do
not reproduce that global fit here; rather, we take its published
parameter \emph{values} as fixed inputs. The single input dataset is
\begin{equation}\label{eq:app1}
  \sin^2\theta_{12}=0.307,\quad \sin^2\theta_{13}=0.02224,\quad
  \sin^2\theta_{23}=0.454,\quad \delta=232^\circ\,(=1.29\pi),
\end{equation}
\begin{equation}\label{eq:app2}
  \Delta m^2_{21}=7.41\times10^{-5}\;\mathrm{eV}^2,\quad
  \Delta m^2_{31}=+2.505\times10^{-3}\;\mathrm{eV}^2.
\end{equation}

\subsection{From $\sin^2\theta$ to mixing angles}

The angles are recovered by
$\theta_{ij}=\arcsin\sqrt{\sin^2\theta_{ij}}$, giving the numerical
values used throughout:
\begin{equation}\label{eq:apptheta}
  \theta_{12}=33.66^\circ,\quad \theta_{13}=8.58^\circ,\quad
  \theta_{23}=42.36^\circ,\quad \delta=232.0^\circ\,(-0.71\pi)\,.
\end{equation}
These are inserted into the PDG parameterization,
Eq.~(\ref{eq:Vexplicit}), to build the $3\times3$ PMNS matrix.

\subsection{Numerical PMNS matrix}

Substitution yields (entries $\approx 4$ significant decimals,
$i\equiv\sqrt{-1}$)
\begin{equation}\label{eq:appU}
  U_\mathrm{PMNS}\;\simeq\;
  \begin{pmatrix}
    0.8232 & 0.5479 & -0.0918+0.1175\,i\\
   -0.3579+0.0659\,i & 0.6494+0.0439\,i & 0.6663\\
    0.4298+0.0723\,i & -0.5233+0.0481\,i & 0.7307
  \end{pmatrix},
\end{equation}
or, in terms of the moduli $|U_{\alpha i}|$,
\begin{equation}\label{eq:appUabs}
  |U_\mathrm{PMNS}|\;\simeq\;
  \begin{pmatrix}
    0.8232 & 0.5479 & 0.1491\\
    0.3639 & 0.6509 & 0.6663\\
    0.4358 & 0.5255 & 0.7307
  \end{pmatrix}.
\end{equation}
The unitarity residual $\max|U^\dagger U-\mathbb{I}|$ evaluates to
$3.3\times10^{-16}$ (machine precision), which we use as a numerical
consistency check of the construction. The Jarlskog invariant
following from Eq.~(\ref{eq:appU}) is $J_\mathrm{CP}\simeq -0.026$,
the value quoted in Eq.~(\ref{eq:J}).

\end{document}